\documentclass[12pt]{article}

\usepackage{graphicx}

\begin{document}
\begin{center}
{\Large\bf \boldmath Monopoles in lattice Electroweak theory} 

\vspace*{6mm}
{B.L.G.~Bakker$^a$, A.I.Veselov$^b$, M.A.~Zubkov$^b$ }\\      
{\small \it $^a$ Department of Physics and Astronomy, Vrije Universiteit,
Amsterdam, The Netherlands \\      
            $^b$ ITEP, B.Cheremushkinskaya 25, Moscow, 117259, Russia}
\end{center}

\vspace*{6mm}

\begin{abstract}
There exist several types of monopole - like topological defects in Electroweak
theory. We investigate properties of these objects using lattice numerical
methods. The intimate connection between them and the dynamics of the theory is
established.
 We find that the density of Nambu monopoles cannot be predicted by the
choice of the initial parameters of Electroweak theory and should be considered
as the new external parameter of the theory. We also investigate the difference
between the versions of Electroweak theory with the gauge groups $SU(2)\otimes
U(1)$ and $SU(2)\otimes U(1)/Z_2$. We do not detect any difference at $\alpha
\sim \frac{1}{128}$. However, such a difference appears in the unphysical
region of large coupling constant $\alpha > 0.1$.
\end{abstract}

\vspace*{6mm}

{
 In both cases we use the following
lattice variables: 1. The gauge field ${\cal U} = (U, \theta)$, where $ U
 \in SU(2), \quad e^{i\theta} \in U(1)$ realized as link variables. 2. A scalar doublet $\Phi_{\alpha}, \;\alpha = 1,2.$
The potential for the scalar field is considered in its simplest form
 in the London limit, i.e., in the limit of infinite bare Higgs
mass. From the very beginning we fix the unitary gauge $\Phi_1 =
\sqrt{\gamma}$, $\Phi_2 = 0$. For the case of the $SU(2) \times U(1)/Z_2$
symmetric model we chose the action of the form (A) $
 S_g  =  \beta \!\! \sum_{\rm plaquettes}\!\!
 ((1-\mbox{${\small \frac{1}{2}}$} \, {\rm Tr}\, U_p \cos \theta_p)
 + \mbox{${\small \frac{1}{2}}$} (1-\cos 2\theta_p))
  + \gamma \sum_{xy}(1 - Re(U^{11}_{xy} e^{i\theta_{xy}}))$,
where the plaquette variables are defined as $U_p = U_{xy} U_{yz} U_{wz}^*
U_{xw}^*$, and $\theta_p = \theta_{xy} + \theta_{yz} - \theta_{wz} -
\theta_{xw}$ for the plaquette composed of the vertices $x,y,z,w$. For the case
of the conventional $SU(2) \times U(1)$ symmetric model we use the action (B)
$S_g  =  \beta \!\! \sum_{\rm plaquettes}\!\!
 ((1-\mbox{${\small \frac{1}{2}}$} \, {\rm Tr}\, U_p )
 + 3 (1-\cos \theta_p))
  + \gamma \sum_{xy}(1 - Re(U^{11}_{xy} e^{i\theta_{xy}}))$.
The following variables are considered as creating a $Z$ boson and a $W$ boson,
respectively: $ Z_{xy}  =  Z^{\mu}_{x} \;
 = {\rm sin} \,[{\rm Arg} U_{xy}^{11} + \theta_{xy}],
 W_{xy}  =  W^{\mu}_{x} \,= \,U_{xy}^{12} e^{-i\theta_{xy}}.$
Here, $\mu$ represents the direction $(xy)$. In the unitary gauge there is also
a $U(1)$ lattice gauge field, which is defined as $ A_{xy}  =  A^{\mu}_{x} \;
 = \,[-{\rm Arg} U_{xy}^{11} + \theta_{xy}]  \,{\rm mod} \,2\pi.$
}
\begin{figure}[h]
\centerline{\includegraphics[width=7cm]{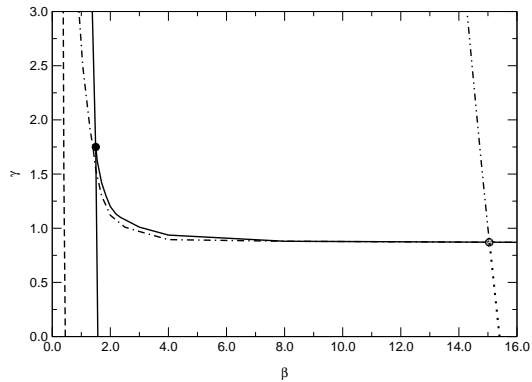}} \caption{The phase diagrams
of the models in the
 $(\beta, \gamma)$-plane.}
\end{figure}
{ The phase diagrams of the two models under consideration are presented in
figure $1$.  The dashed vertical line represents the phase transition in the
$SU(2)\otimes U(1)$-symmetric model. This is the confinement-deconfinement
phase transition corresponding to the $U(1)$ constituents of the model. The
same transition for the $SU(2)\otimes U(1)/Z_2$-symmetric model  is represented
by the solid vertical line. The dashed horizontal line corresponds to the
transition between the broken and symmetric phases of model A. The continuous
horizontal line represents the same transition in model B. Interestingly, in
the $SU(2)\otimes U(1)/Z_2$ model both transition lines meet, forming a triple
point. Real physics is commonly believed to be achieved within the phases of
the two models situated in the right upper corner of Fig.~$1$. The
double-dotted-dashed vertical line on the right-hand side of the diagram
represents the line, where the renormalized $\alpha$ is constant and equal to
$1/128$. All simulations were performed on lattices of sizes $8^4$ and $16^4$.
Several points were checked using a lattice $24^4$. In general we found no
significant difference between the mentioned lattice sizes.

We perform the calculation of renormalized fine structure constant $\alpha_R$
using the potential for infinitely heavy external fermions. We consider Wilson
loops for the right-handed external leptons: $
 {\cal W}^{\rm R}_{\rm lept}(l)  =
 \langle {\rm Re} \,\Pi_{(xy) \in l} e^{2i\theta_{xy}}\rangle.$
Here $l$ denotes a closed contour on the lattice. We consider the following
quantity constructed from the rectangular Wilson loop of size $r\times t$:
${\cal V}(r) = \lim_{t \rightarrow \infty}
 \rm log \frac{  {\cal W}(r\times t)}{{\cal W}(r\times (t+1))}.
$
Due to exchange by virtual photons at large enough distances we expect the
appearance of the Coulomb interaction
$
 {\cal V}(r) = -\frac{\alpha_R}{r} + const.
$

The worldlines of the quantum Nambu monopoles could be extracted from the field
configurations as follows: $ j_Z = \delta \Sigma = \frac{1}{2\pi} {}^*d([d
Z^{\prime}]{\rm mod}2\pi)$
(The notations of differential forms on the lattice  are used here.) The
monopole density is defined as $
 \rho = \left\langle \frac{\sum_{\rm links}|j_{\rm link}|}{4L^4}
 \right\rangle,
$
where $L$ is the lattice size. In order to investigate the condensation of
monopoles we use the percolation probability $\Pi(A)$. It is the probability
that two infinitely distant points are connected by a monopole cluster. We show
Nambu monopole density and percolation probability as a function of $\gamma$
along the line of constant renormalized $\alpha_R = 1/128$. It is clear that
the percolation probability is the order parameter of the transition from the
symmetric to the broken phase \footnote{The hypothesis that Nambu monopole
condensation accompanies the Electroweak transition was first suggested in
\cite{Chernodub_Nambu}. Z-vortices condensation in 3D lattice model of high
temperature Electroweak theory was investigated numerically in
\cite{Chernodub}.}.
\begin{figure}[h]
\centerline{\includegraphics[width=7cm]{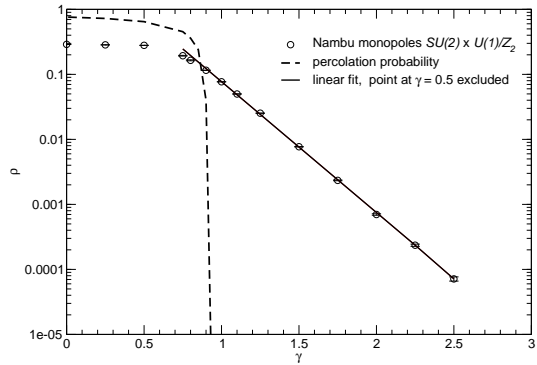}} \caption{Nambu monopole
density and percolation probability as a function of $\gamma$ along the line of
constant $1/\alpha_R=128$.}
\end{figure}
We also measure the magnetic energy (both $SU(2)$ and $U(1)$), which is carried
by Nambu monopoles. The behavior of $\Delta S_p$ shows that a quantum Nambu
monopole may indeed be considered as a physical object.

 In order to evaluate the mass of
the $Z$-boson  we use the zero - momentum correlator: $
\sum_{\bar{x},\bar{y}}\langle \sum_{\mu} Z^{\mu}_{x} Z^{\mu}_{y} \rangle \sim
  e^{-M_{Z}|x_0-y_0|} + e^{-M_{Z}(L - |x_0-y_0|)}$. Here the summation $\sum_{\bar{x},\bar{y}}$ is over the three ``space"
components of the four - vectors $x$ and $y$ while $x_0, y_0$ denote their
``time" components. $L$ is the lattice length in the ``time" direction.
 The physical scale is given in our lattice theory by the value of the
$Z$-boson mass $M^{phys}_Z \sim 91$ GeV. Therefore the lattice spacing is
evaluated to be $a \sim [91 {\rm GeV}]^{-1} M_Z$, where $M_Z$ is the $Z$ boson
mass in lattice units. The real continuum physics should be approached along
the the line of constant $\alpha_R = \frac{1}{128}$, i.e. along the line of
constant physics. We investigated the dependence of the ultraviolet cutoff
$\Lambda = a^{-1} = (91~{\rm GeV})/M_Z$ on $\gamma$ along the line of constant
physics. It occurs that $\Lambda$ is increasing slowly along this line with
decreasing $\gamma$ and achieves the value $430\pm 40$ GeV at the transition
point between the physical Higgs phase and the symmetric phase. According to
our results this value does not depend on the lattice size. This means that the
largest achievable value of the ultraviolet cutoff is equal to $430 \pm 40$ GeV
if the potential for the Higgs field is considered in the London limit.

Our lattice study also demonstrates another peculiar feature of Electroweak
theory. If we are moving along the line of constant $\alpha=1/128$, then the
Nambu-monopole density decreases with increasing $\gamma$ (for $\gamma > 1$).
Its behavior is approximated with a nice accuracy by the simple formula: $\rho
\sim e^{2.08 - 4.6 \gamma}.$ This means that the density of Nambu monopoles in
the continuum theory cannot be predicted by the choice of the usual parameters
of the Electroweak theory and should be considered as a new external parameter
of the theory.

We found that the two definitions of the theory (with the gauge groups
$SU(2)\otimes U(1)/Z_2$ and $SU(2)\otimes U(1)$, respectively) do not lead to
different predictions at the values of $\alpha$ around $1/128$. However, the
corresponding models behave differently at unphysically large values of $\alpha
> 0.1$.  The main difference is in the behavior of the so-called
hypercharge monopoles. }

This work was partly supported by RFBR grants 08-02-00661, and 07-02-00237,
RFBR-DFG grant 06-02-04010, by Grant for leading scientific schools 679.2008.2,
by Federal Program of the Russian Ministry of Industry, Science and Technology
No 40.052.1.1.1112.


\begin{thebibliography}{99}\itemsep -1mm

\bibitem{Nambu}
Y.~Nambu, Nucl.Phys. B {\bf 130}, 505 (1977);\\
Ana~Achucarro and Tanmay~Vachaspati, Phys. Rept. {\bf 327}, 347 (2000); Phys.
Rept. {\bf 327}, 427 (2000).

\bibitem{Chernodub}
M.N.~Chernodub, F.V.~Gubarev, E.M.~Ilgenfritz, and A.~Schiller,
Phys. Lett. B {\bf 434}, 83 (1998);\\
M.N. Chernodub,~F.V. Gubarev, E.M.~Ilgenfritz, and A.~Schiller, Phys. Lett. B
{\bf 443}, 244 (1998).

\bibitem{Chernodub_Nambu} M.N.~Chernodub, JETP Lett. {\bf 66}, 605 (1997)


\bibitem{BVZ2003}
B.L.G.~Bakker, A.I.~Veselov, and M.A.~Zubkov, Phys. Lett. B {\bf  583}, 379
(2004);

\bibitem{BVZ2004}
B.L.G.~Bakker, A.I.~Veselov, and M.A.~Zubkov, Yad. Fiz. {\bf 68}, 1045 (2005).

\bibitem{BVZ2005}
B.L.G.~Bakker, A.I.~Veselov, and M.A.~Zubkov, Phys. Lett. B {\bf 620}, 156
(2005)

\bibitem{BVZ2006}
B.L.G.~Bakker, A.I.~Veselov, and M.A.~Zubkov, Phys. Lett. B {\bf 642}, 147
(2006).


\bibitem{forms}
M.I.~Polikarpov, U.J.~Wiese, and M.A.~Zubkov, Phys. Lett. B {\bf 309}, 133
(1993).


\bibitem{BVZ1999}
B.L.G.~Bakker, A.I.~Veselov, and M.A.~Zubkov, Phys. Lett. B {\bf 471}, 214
(1999).




\bibitem{Z2007}
M.A.~Zubkov, Phys. Lett. B {\bf  649}, 91 (2007).




\bibitem{12}
W.Langguth, I.Montvay, P.Weisz Nucl.Phys.B277:11,1986.

\bibitem{13}
W. Langguth, I. Montvay (DESY) Z.Phys.C36:725,1987

\bibitem{14}
Anna Hasenfratz, Thomas Neuhaus, Nucl.Phys.B297:205,1988


\end{thebibliography}
\end{document}